\documentclass[twocolumn,prl,aps,amsfonts]{revtex4}
\usepackage{graphicx}
\usepackage{psfrag, color}
\usepackage{verbatim}

\begin{document}

\title{Transport and percolation in a low-density
high-mobility two-dimensional hole system}

\author{M. J. Manfra$^1$, E. H. Hwang$^2$, S. Das Sarma$^2$, 
  L. N. Pfeiffer$^1$, K. W. West$^1$, and A. M. Sergent$^1$}

\affiliation{${}^1$ Bell Laboratories, Alcatel-Lucent, Murray Hill, New
  Jersey, 07974 
\\ ${}^2$ Condensed Matter Theory Center, Department of Physics,
University of Maryland, College Park, Maryland, 20742}

\begin{abstract}
We present a study of the temperature and density dependence of the
resistivity of an extremely high quality
two-dimensional hole system grown on the (100) surface of GaAs.  For
high densities in the metallic regime ($p\agt 4 \times 10^{9}$ cm$^{-2}$),
the nonmonotonic temperature dependence ($\sim 50-300$ mK) of the
resistivity is 
consistent with temperature dependent screening of residual
impurities.  At a fixed temperature of $T$= 50 mK, the conductivity
vs. density data indicates an inhomogeneity driven percolation-type
transition to an insulating state at a critical density of
$3.8\times 10^9$ cm$^{-2}$.
\end{abstract}

\maketitle
The low temperature conduction properties of an interacting
two-dimensional (2D) electron gas in the presence of disorder remains
a topic of fundamental interest and some controversy.  It has been
understood for some time that all non-interacting electronic states
are localized in the presence of disorder at $T$=0 \cite{Abrahams}. This model was
challenged by the observation of Kravchenko and co-workers
\cite{Kravchenko} of an apparent metal-to-insulator transition in Si
MOSFETs.  Subsequently a variety of models have been put forth
\cite{review,review1,Punnoose,Zala} to
explain a wealth of experimental data.  The 2D metal-to-insulator
transition (MIT) has now been studied in a variety of semiconductor systems \cite{Kravchenko, Lai1, Lilly, Zhu, Anissimova, DasSarma1, Simmons, Yoon, Allison}.  Yet the question remains: does the transition from insulating
to metallic behavior observed in certain 2D systems represent a
quantum phase transition to a novel metallic state stabilized by
interactions \cite{Punnoose} or can
the experimental data in the metallic regime be explained in terms of finite-temperature Fermi liquid
behavior \cite{review1}?  
The MIT in Si MOSFETs has been interpreted as
evidence for quantum critical behavior \cite{Anissimova}.  However, recent experimental work
\cite{DasSarma1,Tracy} in ultra-high mobility n-GaAs 2D systems has
interpreted the MIT as a density inhomogeneity induced percolation transition.

In this Letter, we present a study of the transport
properties of a clean two-dimensional hole system (2DHS) in the
vicinity of the putative metal-to-insulator transition.  Our
carbon-doped 2DHSs display mobilities in excess of
2$\times$10$^6$ cm$^2$/Vs at densities
p$\leq$3$\times$10$^{10}$  cm$^{-2}$, the highest ever achieved in a
2DHS in this density range.  The use of extremely high mobility
samples allows us to probe the regime of large $r_s$ ($r_s$=E$_C$/E$_F$, where E$_C$ is the Coulomb interaction scale and E$_F$ is the Fermi energy) in a clean
system, where the effects of carrier-carrier interactions should be
most apparent.
Interestingly, we find that for high densities in the metallic regime the nonmonotonic temperature
dependence of the resistivity is consistent with temperature dependent
screening of residual impurities.  At a fixed temperature of $T$=50 mK,
the conductivity vs. density data indicates a disorder-induced
percolation-type transition to an insulating state at low density.   While the critical density ($p_c$ $\sim 3.8\times 10^9$ cm$^{-2}$) and
the interaction parameter ($r_s \sim 45$) for our sample are among the
lowest and highest ever respectively recorded in the 2D MIT
literature, we find that the data are well explained within this simple framework.  The fact that our critical density is similar to the critical
density for 2D MIT found in ref. \cite{DasSarma1} suggests
that the effective 2D MIT in GaAs electron and hole systems at experimentally accessible temperatures is a 
disorder-driven crossover phenomenon and not an
interaction driven quantum phase transition, since the $r_s$ value for our hole system is four times
larger than the corresponding $r_s$ in the electron system of
ref. \cite{DasSarma1}.

Our experiments utilize recently developed structures in which the
2DHS resides in a 20nm GaAs/AlGaAs quantum well grown on the (100)
surface of GaAs that has been modulation-doped with carbon
\cite{manfra}.  A back-gated Hall bar
geometry is used in this experiment to tune the density from
2.9$\times$10$^{10}$ cm$^{-2}$ to 2.9$\times$10$^{9}$ cm$^{-2}$ in
single sample.  Cyclotron resonance studies of similarly grown
carbon-doped quantum wells indicate that the hole effective 
mass is large, $m_{h}\sim 0.3m_e$, where $m_e$ is the free electron
mass \cite{Han}.  To give a comparison with other 2D systems studied in the literature,
we mention that the 2D n-GaAs samples of refs. \cite{Lilly,DasSarma1}
have an electron mobility of $5\times10^6$ cm$^2$/Vs at the carrier
density of $3\times10^{10}$ cm$^{-2}$ compared with our hole mobility
of $2\times 10^6$ cm$^2$/Vs at the same density. Taking into account the effective mass difference ($m=0.07m_e$
for electrons in refs. \cite{Lilly} and \cite{DasSarma1} and
$m\approx 0.3m_e$ for holes in our samples) and remembering that
mobility $\mu \propto \tau/m$, where $\tau$ is the transport
relaxation time, we conclude that our 2D hole samples are comparable in
quality to the corresponding 2D electron samples
of refs. \cite{Lilly,DasSarma1} while $r_s$ is significantly larger. In this manner the data presented here probe a regime inaccessible to the experiments of refs. \cite{Lilly, DasSarma1}.

Figure 1 displays magnetotransport at $T$=50 mK.  The
exceptional quality of these samples is evident in the strength of the
fractional quantum Hall state at $\nu$=2/3 at B$\sim$1 Tesla and the
nascent features observed at $\nu$=5/3 and 4/3 at even lower values of
magnetic field. Figure 2 displays the temperature dependence of the resistivity at B=0
for several values of density p$\leq$9$\times$10$^{9}$ cm$^{-2}$.
At all but the lowest densities, the temperature
dependence is strongly nonmonotonic \cite{Allen}.  Starting from the high
temperature regime ($T$$\sim$1 K) the resistivity first increases with
decreasing temperature, reaches peak value, and then decreases rapidly
with further reduction in temperature. At p=9.0$\times$10$^9$ cm$^{-2}$, the low temperature drop in resistance is a factor of 2.  As the density is reduced this
crossover point, where $d\rho / dT$ changes sign, moves to lower
temperature, presumably tracking the reduction in the Fermi energy of
our 2D system.  The metallic behavior persists to very
low density, at p=4.1$\times$10$^{9}$ cm$^{-2}$  $d\rho/dT$ still
changes sign near T=50 mK.  At p=2.9$\times$10$^{9}$ cm$^{-2}$, the
lowest density accessed in our study, $\rho$ vs. $T$ is monotonic,
suggesting that the sample is approaching an insulating regime.
Yet at p=2.9$\times$10$^{9}$ cm$^{-2}$ the conductivity is not exponentially
suppressed with decreasing temperature as expected for an insulating
state, but rather, the conductivity displays a linear dependence on
temperature (i.e. $\rho \propto T^{-1}$) in the $T=50-450$ mK range \cite{Nohlinear}.
In this regime of low density, the Fermi temperature is as low as $T_F \sim 250$ mK.  The impact of carrier degeneracy and
the temperature dependence of ionized impurity scattering must be
considered as discussed below.

\begin{figure}
\scalebox{0.75}{\includegraphics{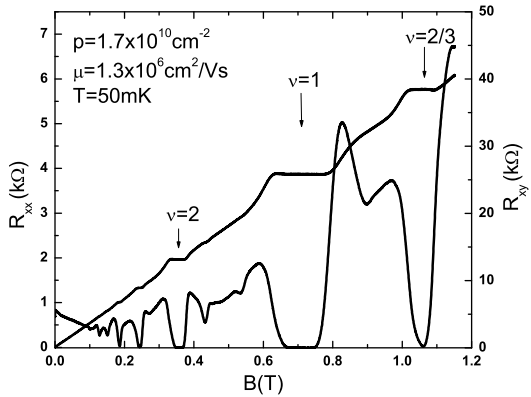}}
\caption{Longitudinal and Hall resistance at zero gate bias.  The
  fractional quantum Hall state at $\nu$=2/3 is fully developed at
  B$\sim$1T while weaker features at $\nu$=5/3 and 4/3 are also
  visible.} 
\end{figure}

\begin{figure}
\scalebox{0.75}{\includegraphics[width=\columnwidth]{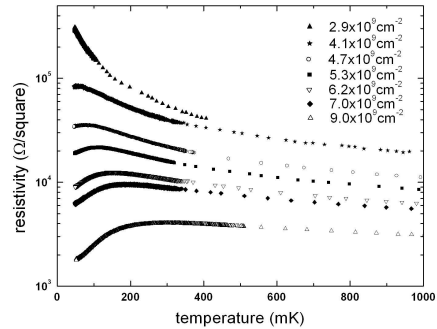}}
 \caption{Temperature dependence of the resistivity for
   50 mK$\leq$T$\leq$1 K for densities ranging from 9.0x10$^9$ cm$^{-2}$
   to 2.9x10$^9$ cm$^{-2}$.} 
\label{fig_rho}
\end{figure}
\begin{figure}
\scalebox{0.75}{\includegraphics[width=\columnwidth]{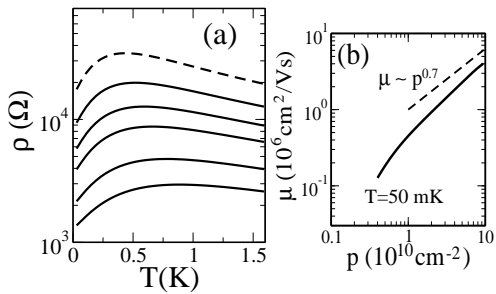}}
\caption{
(a) Calculated resistivities
for p-GaAs quantum well system ($a=200$ \AA) for various hole
densities $p=$0.4, 0.5, 0.6, 0.7, 0.9, 1.1 $\times
10^{10}$ cm$^{-2}$ (from the top) with linearly screened charge
impurity scattering. (The dashed curve is in the $k_F l <1$ regime.)
(b) Calculated mobility as a function of hole density at $T$=50 mK.
The dashed straight line indicates the $\mu \sim p^{0.7}$ behavior. 
We only show mobility in the $k_Fl>1$ regime.
}
\label{fig_th3}
\end{figure}
\begin{figure}
\scalebox{0.75}{\includegraphics{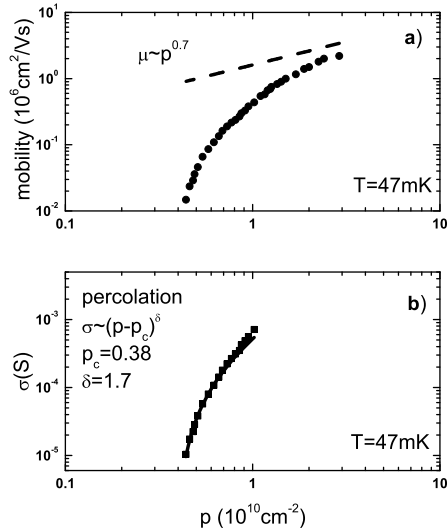}}
 \caption{(a) Mobility vs. density at fixed temperature $T$=47 mK.  (b)
   Conductivity vs. density (solid squares) along with the fit
   generated assuming a percolation transition. The dashed line in (a)
 indicates the $\mu \sim p^{0.7}$ behavior.} 
\label{mobility}
\end{figure}

In order to understand the low temperature 2D transport properties in the metallic regime
we have carried out a microscopic transport calculation using the
Boltzmann theory \cite{review1}. In our calculation we assume
that the 2D carrier resistivity is entirely limited by screened
impurity scattering. We neglect all phonon scattering effects 
because our theoretical
estimate shows phonon scattering to be negligible for 2D holes in
GaAs structures in the $T<1 K$ regime of interest to us.  Formally,
this approach is well-controlled in the regime of small $r_s$ and
$k_Fl\gg1$, where $k_F$ is the Fermi wavevector and $l$ is the mean
free path. While our Fermi liquid approach is not well controlled at high interaction, it has been proven to work at relatively large $r_s $ which substantiates its use in the high density, large conductance regime $k_Fl > 1$.
We therefore mostly show our theoretical results in the $k_F l>1$
metallic regime except for the lowest density (dashed) curve in
Fig. 3(a). We emphasize that by adjusting the charged impurity density
and by including disorder effects in the screening function, it is possible to
get reasonable quantitative agreement between our Boltzmann-Fermi
liquid theory [Fig. 3(a)] and the experimental data (Fig. 2) even in
the insulating low-density $k_F l <1$ regime \cite{sds_in}.  We show our calculated results in Fig. \ref{fig_th3} for various hole
densities. In our calculation the overall resistivity scale depends on
the unknown charge impurity density, but the qualitative trends in
$\rho(T,p)$ arise from basic aspects of the underlying scattering
mechanism.
Our calculated resistivity, Fig. \ref{fig_th3}(a), shows a non-monotonicity
(as seen in Fig. \ref{fig_rho} of the experimental data) where it
increases with T at first and then decreases 
with increasing T after going through a  maximum at a
density-dependent temperature $T_m$ where $d\rho/dT$ changes sign. 
This non-monotonicity is a non-asymptotic
finite temperature ``quantum-classical'' crossover phenomenon, and
arises from the competition between  
two contributions to transport \cite{review1}, i.e. the
temperature dependence of the dielectric screening function and
the thermal averaging of the scattering rate. The screening effect 
decreasing with increasing temperature gives rise to
increasing effective disorder with increasing T , and hence
increasing $\rho(T)$ with T. Based on the Boltzmann theory, however,
in the higher temperature classical non-degenerate regime 
the thermal averaging of the scattering rate 
becomes quantitatively more important and shows stronger temperature
dependence than screening, which gives
rise eventually to a $\rho(T)$ decreasing with T for $T > T_m$. Thus,
if the system is homogeneous and not strongly localized this crossover behavior
appears in the transport measurements as can be seen in
Fig. \ref{fig_rho}.  We note that as the density is reduced significantly below $p=9.0\times 10^{9}$ cm$^{-2}$, $k_Fl\sim 1$ and direct comparison between theory and experment becomes more problematic.  Nevertheless, we point out that the theory reproduces well all the qualitative
features of the data in the metallic regime shown in Fig. \ref{fig_rho}: (1)
An approximately (and upto) a factor of 2 increase in low-temperature
$\rho(T)$ with increasing $T$; (2) a maximum in $\rho(T)$ at a density
dependent $T_m(p)$; (3) a ``high-temperature'' decrease of $\rho(T)
\sim T^{-1}$ for $T>T_m$.  One important difference between the 2D n-GaAs electron system
\cite{Zhu,Lilly,DasSarma1} and our 2D p-GaAs hole system is the very
strong metallic temperature dependence ($\sim$ upto 200\%) we observe
(cf. Fig. 2) in our hole sample versus the rather weak ($\sim$ 20\%
only) temperature dependence in the metallic phase of the n-GaAs
electron system \cite{Lilly,DasSarma1}. This strong (weak) metallic
temperature dependence in the 2D hole (electron) GaAs system is easily
explained by the screening theory \cite{DasSarma2} as arising from the
much larger (by a factor four) dimensionless screening parameter
$q_{TF}/2k_F$ in the hole system compared with the electron system due
to the larger hole effective mass.
In Fig. \ref{fig_th3}(b) we show the calculated mobility as a function of hole
density at $T=50$ mK. At high densities ($p \ge 1.0 \times 10^{10}$ $cm^{-2}$) the exponent $\alpha$ in $\mu \propto p^{\alpha}$ is about 
0.7, which is consistent with experiment (cf. Fig. \ref{mobility})
indicating that the screened 
charged impurity is the main scattering source.  

\begin{figure}
\scalebox{0.75}{\includegraphics{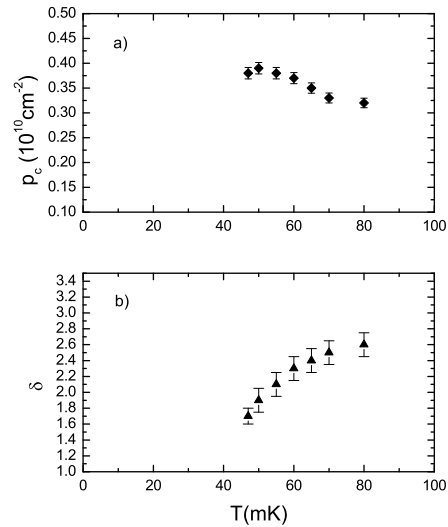}}
 \caption{a) Variation of the derived critical density p$_c$ as a
   function of temperature.  b) Variation of $\delta$ with
   temperature.} 
\end{figure}

In order to gain
further insight into the transition to insulating behavior at low density we now
focus on the density dependence of the conduction properties at fixed
low temperature. 
Figure 4(a) shows the dependence of mobility on density at fixed
temperature T=47 mK.  While the data for all densities cannot be
described by a single 
power law behavior, at densities above
$\sim$2$\times$10$^{10}$ cm$^{-2}$ the mobility approaches an approximate
power law behavior $\mu \sim p^{0.7}$ as seen in our theory. At low
densities, p$\leq$ 1 $\times$ 10$^{10}$ cm$^{-2}$, the power law
changes to a much stronger function of density, pointing to a change
in the screening properties of the system at low carrier density. We emphasize that the quantitative and qualitative discrepencies between the lowest density data of Fig. 4a and the calculation of Fig. 3b indicate that the screening theory which captures the essential experimental features at higher densities does not fully describe the transition to insulating behavior.

For high mobility n-type GaAs, Das Sarma {\it et al.} \cite{DasSarma1}
interpreted the density driven crossover from metallic 
conductivity to an insulating regime as a percolation-type transition.  
The importance of interactions as parameterized by
$r_s$ for the MIT in GaAs remains an open
question.  In high mobility n-type 2D layers, the regime of
$r_s$$\sim$10 can be reached at low density.  Due to the higher
effective mass in our 2DHS, $r_s$ reaches approximately 45 at low
density in this study.  It is an interesting question if a percolation
type analysis can also be applied to our ultra high mobility 2DHS in
the regime of large $r_s$.  
We mention that there have also been various percolation analyses of
2D MIT in other n- and p-GaAs systems \cite{Allison,Leturcq} for low mobility
(a factor of 10 lower than our mobility) systems where the temperature
dependent metallic behavior is much weaker than what we see in our
high-mobility system.

To describe the transition of the resistivity from metallic behavior
to an apparent insulating behavior at lowest densities, we fit our
data using a percolation transition model \cite{percolation}. The
physical motivation for 
a percolation transition is readily apparent.  As the density of the
2DHS is reduced, the system can no longer fully screen the random
disorder potential.  The 2DHS breaks up into ``puddles'' of carriers
isolated from one another by peaks in the disorder potential.
Transport then takes place through percolating conduction paths in the
disorder potential network.  In the percolation model, the
conductivity near the percolation threshold density is described by: 
$$\sigma(p)=A(p-p_c)^{\delta}$$
where $\sigma(p)$ is the density dependent conductivity, $p_c$ is the
critical density, and $\delta$ is the critical exponent descibing the
transition.  $A$ is a constant of proportionality.  The 2D percolation
exponent is expected to be 1.31 \cite{DasSarma1,Leturcq,percolation}.
The results of such an analysis applied to our sample is shown in Figure 4(b).  The
extracted exponent for our sample is $\delta$=1.7$\pm$0.1 and the
critical density
$p_c$=0.38$\times$10$^{10}$ cm$^{-2}$$\pm$0.01$\times$10$^{10}$ cm$^{-2}$. Formally, the preceding analysis is valid for $(p-p_c)\ll p_c$.  While the quality of the fit does not depend sensitively on the number of data points included, clearly the restriction $(p-p_c)\ll p_c$ must be relaxed somewhat to make connection with the experimental data.
Interestingly, the extracted conductivity parameters from the
percolation analysis are consistent with known behavior of very high
mobility n-type GaAs 2D systems as reported in ref. \cite{DasSarma1}.
In particular, our extracted percolation exponent of 1.7 at $T$=47 mK is
almost identical with the corresponding exponent value at a somewhat
higher temperature ($\sim 100-200$ mK) in ref. \cite{DasSarma1}, which is
consistent with equivalent values of the dimensionless temperature
$T/T_F$ for the two systems, where $T_F$ is the Fermi temperature, and
our critical density 
of $3-4\times 10^9$ cm$^{-2}$  is consistent with the corresponding critical
electron density in ref. \cite{DasSarma1}. The temperature dependences of the fit parameters are detailed in figure 5.
To conclude, we find
that the temperature dependence of the metallic phase is
well-described by assuming the conductivity to be limited by
temperature and density dependent screened charged impurity scattering
whereas the apparent low-density ($\sim 3.8\times 10^9$ cm$^{-2}$)
transition from the metal to the insulating phase is well-described by
a density-inhomogeneity driven percolation transition.

\end{document}